\documentclass[prb,amsfonts,amssymb,floats,twocolumn,aps]{revtex4-2}

\usepackage[final,dvips]{epsfig}
\usepackage{amsmath,amsfonts,amssymb}
\usepackage[T1]{fontenc}
\usepackage{bbm}
\usepackage{float}
\usepackage[caption = false]{subfig}
\usepackage{graphicx}
\usepackage{color}
\usepackage{hyperref}
\usepackage[normalem]{ulem}
\usepackage{braket}

\newcommand{\ai}{a}
\newcommand{\ad}{a^{\dagger}}
\newcommand{\bi}{b}
\newcommand{\bd}{b^{\dagger}}
\newcommand{\kk}{\vec{k}}
\newcommand{\ci}{c}
\newcommand{\cd}{c^{\dagger}}
\newcommand{\di}{d}
\newcommand{\dd}{d^{\dagger}}
\newcommand{\gk}{\gamma_{\kk}}
\newcommand{\gka}{\gamma^{a}_{\kk}}
\newcommand{\gkb}{\gamma^{b}_{\kk}}

\begin{document}
\title{$J_{1} - J_{2} - \delta$ model on a square lattice: From Altermagnet to Columnar antiferromagnet via quantum disordered phase}


\author{Subharthi Paul}
\affiliation{Tata Institute of Fundamental Research, Hyderabad 500046, India}

\author{Darshan G. Joshi}
\affiliation{Tata Institute of Fundamental Research, Hyderabad 500046, India}

\begin{abstract}
Altermagnetism in the case of local-moment magnets is characterized by zero net magnetization and splitting of opposite chirality magnon bands in the absence of any spin-anisotropic interactions or external field.
In this work we investigate the robustness of altermagnets against quantum fluctuations arising from magnetic frustration. 
We consider a Heisenberg model on a square lattice with two different types of second-neighbor interactions, as in a checkboard pattern, in addition to the nearest-neighbor interaction. This model continuously interpolates between the $J_{1}-J_{2}$ Heisenberg model on the square lattice and the Heisenberg model on the checkerboard lattice. For weaker second-neighbor interactions a N\'{e}el-type altermagnet phase is realized. On the other hand, for stronger second-neighbor interactions a columnar antiferromagnet emerges.
Using linear spin-wave theory we calculate the magnon dispersion, order parameter, static and dynamical structure factors for the two magnetic phases. Further, we show that there is an intermediate quantum disordered phase separating the two magnetic phases, which is connected to the one realized in the $J_{1}-J_{2}$ model. The quantum disordered region is further stabilized as we tune towards the checkerboard lattice limit. 
\end{abstract}

\maketitle 

\section{Introduction}
\label{sec:intro}

Magnetic frustration often leads to strong quantum fluctuations, resulting in the destruction of conventional magnetic phases in favor of quantum disordered phases \cite{Vojta_2018}. 
A classically degenerate ground-state energy manifold resulting from frustration in a local-moment system is often the source of exotic quantum phases, such as quantum spin liquids. One of the paradigmatic models where this has been long studied is the $J_{1} - J_{2}$ Heisenberg antiferromagnetic model on a square lattice. 
Quantum fluctuations destroy the classical phases in the vicinity of the fully frustrated point, $J_{2}/J_{1} = 0.5$, which has a large classical degeneracy. This was first shown using linear spin-wave theory \cite{Chandra_swt}. Subsequently, this model has attracted a lot of attention due to the possibility of a quantum spin-liquid phase in this quantum disordered region, apart from certain valence-bond solid phases \cite{Chubukov91, Zhitomirsky96, Richter10, Jiang12, Gong14, Wang18, Ferrari20}. Appearance of such an intermediate quantum disordered phase with possible quantum liquid, apart from magnetic long-range order, is also seen in other frustrated quantum models such as the $J_{1}-J_{2}$ Heisenberg model on the triangular lattice \cite{Jolicoeur89, Zhu15, Iqbal16} and Heisenberg model on a frustrated cubic lattice \cite{Laubach16}.

Apart from quantum disordered phases, magnetic frustration may also lead to novel magnetic phases \cite{Vojta_2018}. One such recent example is that of altermagnets \cite{Smejkal1, Smejkal2}, where in the case of electronic systems the electron bands exhibit spin splitting in certain regions of the momentum space, while the net magnetization is still zero. Altermagnets, in the context of local moments, exhibit  splitting of the magnon bands even in the absence of any spin-anisotropic interactions. Furthermore, these split magnon bands have an associated chirality \cite{Consoli_2025}. One of the minimal spin models to study altermagnet is the Heisenberg antiferromagnet on a checkerboard lattice \cite{Consoli_2025}. In fact, this model was studied long ago, and magnon splitting was already reported there \cite{checker_swt}. 

From a fundamental perspective, it is important to understand the role of quantum fluctuations in altermagnets and possible quantum phase transitions associated with it. An interesting possibility is the  emergence of a quantum disordered phase, particularly a quantum spin liquid, from an altermagnet. It is therefore imperative to go beyond the scope of minimal model of altermagnets. 
In this work, we consider a frustrated lattice model that hosts an altermagnet as well as other conventional magnetic phase. By systematically studying quantum fluctuations within the large-$S$ spin-wave theory, we show that this model hosts a quantum disordered phase in the vicinity of an altermagnet. 

The rest of the paper is organized as follows: In Sec. \ref{sec:model} we introduce our model and terminology. In Sec. \ref{sec:neel} we discuss the altermagnet phase in detail using the linear spin-wave theory, while in Sec. \ref{sec:col} we present our calculations in the columnar phase. We discuss the resulting phase diagram of our model  and give concluding remarks with future prospects in Sec. \ref{sec:con}. The technical details of our calculations are presented in the corresponding appendices.

\section{model}
\label{sec:model}

\begin{figure}
\centering
\subfloat[]{\includegraphics[width=0.33\linewidth]{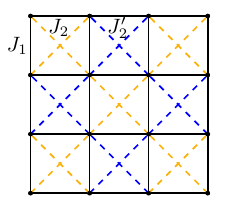}}
\subfloat[]{\includegraphics[width=0.33\linewidth]{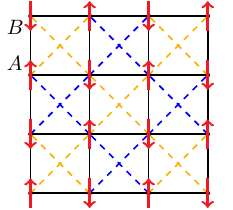}}
\subfloat[]{\includegraphics[width=0.33\linewidth]{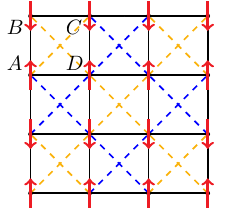}}
\caption{(a) Figure showing the exchange interactions described in model in Eq. (\ref{eq:model}). Black bonds are NN interactions with strength $J_{1}$, while yellow and blue bonds are NNN interactions with strength $J_{2} = \lambda J (1+\delta)$ and $J'_{2} = \lambda J (1-\delta)$ respectively.  (b) Classical spin orientation in N\'{e}el/Altermagnet phase with $A$ and $B$ being the two sublattices. This state is the starting point for the spin-wave expansion for $\lambda \leq 0.5$. (c) Classical spin orientation in the Columnar phase with $A$, $B$, $C$ and $D$ being the four sublattices. This state is the starting point for the spin-wave expansion for $\lambda > 0.5$. }
\label{fig:model}
\end{figure}

We consider a square lattice nearest-neighbor Heisenberg model with additional second neighbor frustrating interactions as shown in Fig. \ref{fig:model}. The Hamiltonian is given by
\begin{equation}
H = J \sum_{\langle ij \rangle} \vec{S}_{i}\cdot \vec{S}_{j} 
+ J_{2} \sum_{\langle\langle ij \rangle\rangle} \vec{S}_{i}\cdot \vec{S}_{j} 
+ J'_{2} \sum_{\langle\langle ij \rangle\rangle'} \vec{S}_{i}\cdot \vec{S}_{j}  \,,
\label{eq:model}    
\end{equation}
where the first term is the nearest-neighbor (NN) Heisenberg interaction with $J>0$. There are two types of second-neighbor (NNN) Heisenberg interactions on alternating square plaquettes (see Fig. \ref{fig:model}). These are the second and third terms in Eq. (\ref{eq:model}) above, with $J_{2} = \lambda J (1+\delta)$ and $J'_{2} = \lambda J (1-\delta)$, shown as yellow and blue bonds in Fig. (\ref{fig:model}) (a), and the NNN summations denoted by $\langle\langle ij \rangle\rangle$ and $\langle\langle ij \rangle\rangle'$ respectively. The parameter $\lambda = (J_{2} + J'_{2})/2J >0$ determines the relative strength of the second-neighbor interactions with respect to $J$, and the parameter $\delta = (J_{2} - J'_{2})/(J_{2} + J'_{2}) \in [0,1]$ parametrizes the difference between the two types of second-neighbor interactions. It is clear that the parameter $\delta$ leads to interpolation between the  $J_{1}-J_{2}$ Heisenberg model on a square-lattice ($\delta=0$) and the Heisenberg model on a checkerboard lattice ($\delta=1$). This model has been recently studied in the context of altermagnets \cite{Brekke2023, alter_nlswt}, for $\lambda < 1/2$. However, a thorough analysis as a function of $\lambda$ and $\delta$ has not been done to our knowledge. Below we first recall the phases of this model studied in specific limits. 

Let us first consider $\delta=0$, where the model in Eq. (\ref{eq:model}) reduces to the well-studied $J_{1}-J_{2}$ Heisenberg model on a square lattice. From spin-wave theory based and other series expansion results, for $\lambda \lesssim 0.4$ the model realizes a N\'{e}el antiferromagnetic phase with an ordering wavevector, $\vec{Q}_{N} = (\pi,\pi)$, while for $\lambda \gtrsim 0.5$ the model has a columnar antiferromagnetic (CAF) phase with ordering wavevector, $\vec{Q}_{C} = (0,\pi)$ or $\vec{Q}'_{C} = (\pi,0)$ \cite{Chandra_swt, Igarashi92, Igarashi93, Oitmaa96, Igarashi2005, Majumdar2010}. The intermediate region around $\lambda \approx 1/2$ is a quantum disordered phase and has been a subject of continued interest in the context of quantum spin liquids (QSLs). Different bond-operator perturbation theory and numerical results indicate the presence of a QSL phase, along with possible other dimer or plaquette phases in this intermediate region around $\lambda \approx 1/2$ \cite{Chubukov91, Zhitomirsky96, Richter10, Jiang12, Gong14, Wang18, Ferrari20}. 

At $\delta=1$ the third term in Eq. (\ref{eq:model}) vanishes, resulting in the Heisenberg model on a checkerboard lattice. This model was previously studied in Ref. \cite{checker_swt} and shown to host antiferromagnetic order for $\lambda \lesssim 0.375$ with a quantum disordered phase for $\lambda \gtrsim 0.375$. In Ref. \cite{checker_swt}, it was already identified that for finite $\lambda$ there is a splitting in the magnon bands. In the more recent terminology, this phase is labeled as an altermagnet (AM) and has been recently studied \cite{Consoli_2025, alter_nlswt}. Nonlinear spin-wave theory calculations in the AM phase for $\delta \in [0,1]$ and $\lambda < 1/2$ has been performed recently \cite{alter_nlswt}.
In this work, we systematically study the model in Eq. (\ref{eq:model}) for different values of $\delta$ and $\lambda$ using the linear spin-wave theory, and plot the phase diagram. As we shall see, the intermediate quantum disordered phase present in the $J_{1}-J_{2}$ Heisenberg model is further stabilized upon increasing $\delta$. 

\section{Altermagnet phase}
\label{sec:neel}

\begin{figure}
\centering
\subfloat[]{\includegraphics[width=0.5\linewidth]{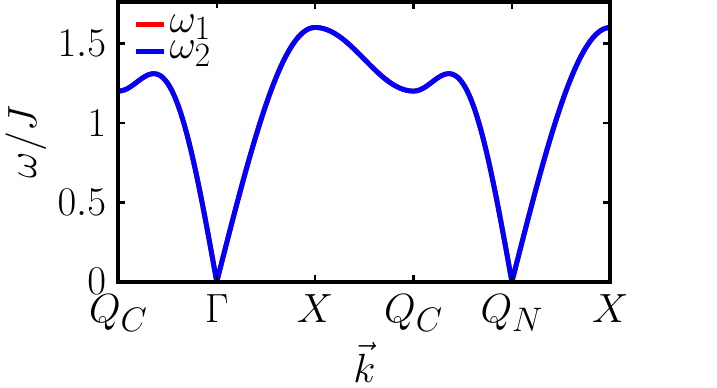}} 
\subfloat[]{\includegraphics[width=0.5\linewidth]{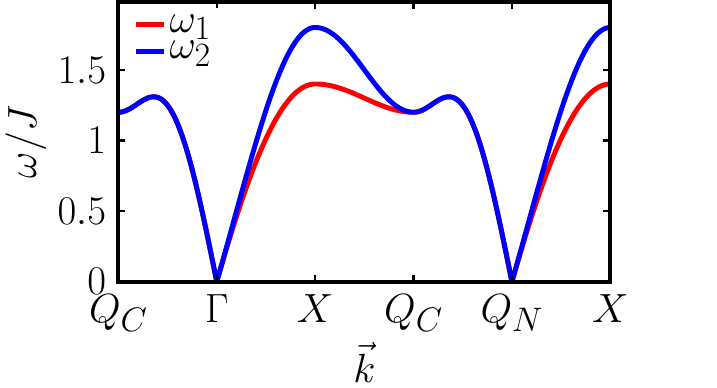}} \\ 
\subfloat[]{\includegraphics[width=0.5\linewidth]{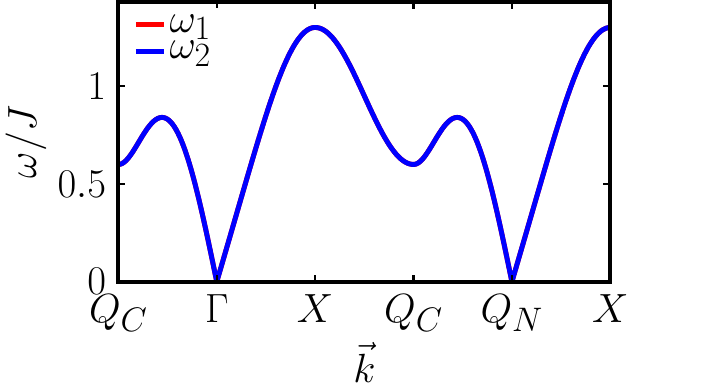}} 
\subfloat[]{\includegraphics[width=0.5\linewidth]{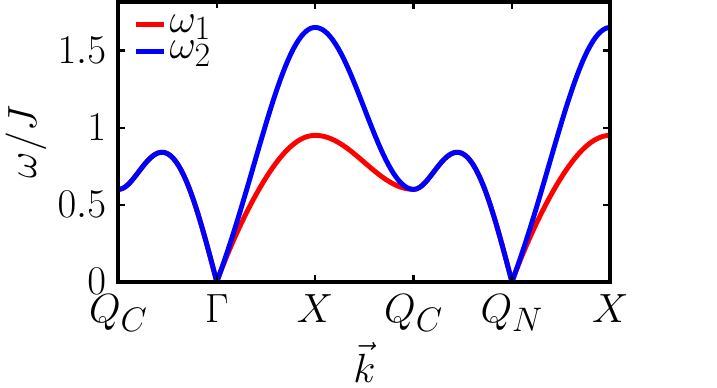}}
\caption{Magnon dispersion in the Altermagnet/N\'{e}el phase. (a) $\delta=0, \lambda=0.2$, (b) $\delta=0.5, \lambda=0.2$, (c) $\delta=0, \lambda=0.35$, (d) $\delta=0.5, \lambda=0.35$. At $\delta=0$, there is no magnon band splitting and the phase is the standard N\'{e}el phase. For any $\delta \neq 0$ there is a splitting of magnon bands, as can be see in plots (b) and (d).
}
\label{fig:wn}
\end{figure}

\begin{figure}
\centering
\subfloat[]{\includegraphics[width=0.25\textwidth]{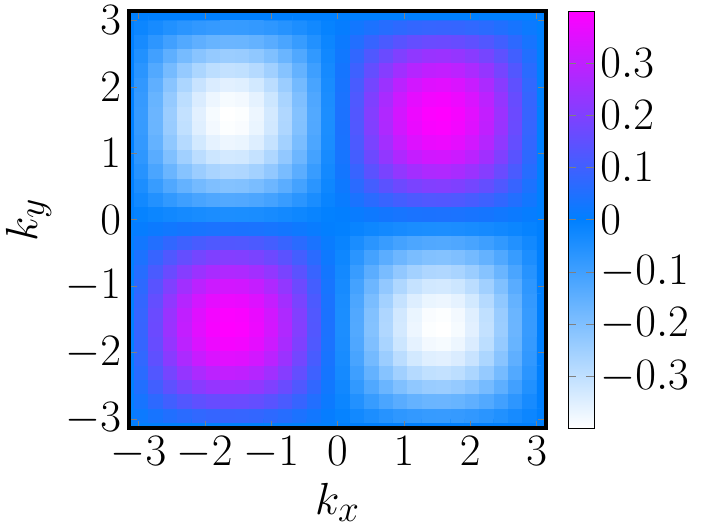}} 
\subfloat[]{\includegraphics[width=0.5\linewidth]{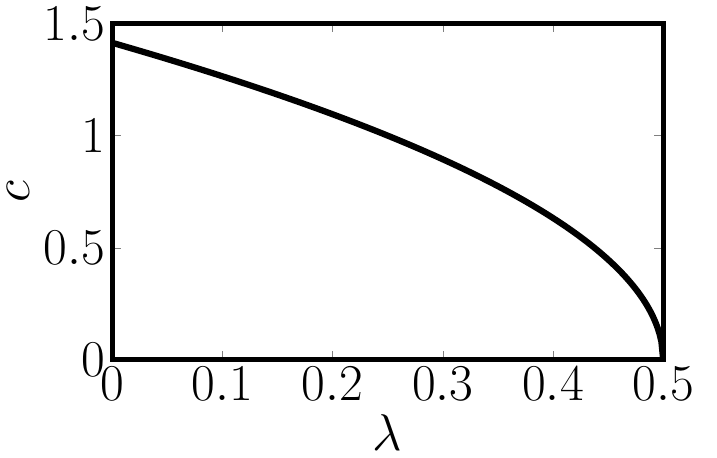}}
\caption{(a) Magnon band splitting, $\delta \omega_{\kk}$, (Eq. \ref{eq:mags}) for $\lambda= 0.2$ and $\delta=0.5$. 
(b) Spin-wave velocity, Eq. (\ref{eq:swcn}), in the AM phase as a function of $\lambda$. As noted in the text, this is independent of $\delta$.
}
\label{fig:mags}
\end{figure}

In this section we discuss the spin-wave theory of the Altermagnet phase realized for $\lambda <1/2$. We first recall the Holstein-Primakoff transformation on the two sublattices (see Fig. (\ref{fig:model} b)  ),
\begin{align}
S^{z}_{Ai} &= S - \ad_{i} \ai_{i} \,, && S^{z}_{Bi} = \bd_{i} \bi_{i} - S \,, \nonumber \\ 
S^{+}_{Ai} &= \sqrt{2S - \ad_{i}\ai_{i}} ~\ai_{i} \,, && S^{+}_{Bi} = \bd_{i} \sqrt{2S - \bd_{i}\bi_{i}}  \,, \nonumber \\ 
S^{-}_{Ai} &= \ad_{i} \sqrt{2S - \ad_{i}\ai_{i}} \,, && S^{-}_{Bi} = \sqrt{2S - \bd_{i}\bi_{i}} ~\bi_{i} \,,
\label{eq:hp}
\end{align}
where $\ai, \bi$ ($\ad, \bd$) are bosonic magnon annihilation (creation) operators on the two sublattices respectively. The transformation is written for an arbitrary value of $S$. Now formally, in the limit of large-$S$, the square-roots in Eq. (\ref{eq:hp}) can be expanded order by order in powers of $1/S$, thus leading to a systematic $1/S$ spin-wave expansion. The leading order terms in this expansion  constitute the linear spin-wave theory (LSWT). To this end,
\begin{align}
S^{z}_{Ai} &= S - \ad_{i} \ai_{i} \,, && S^{z}_{Bi} = \bd_{i} \bi_{i} - S \,, \nonumber \\ 
S^{+}_{Ai} &\approx \sqrt{2S} ~\ai_{i} \,, && S^{+}_{Bi} \approx \sqrt{2S} ~\bd_{i}  \,, \nonumber \\ 
S^{-}_{Ai} &\approx \sqrt{2S} ~\ad_{i} \,, && S^{-}_{Bi} \approx \sqrt{2S} ~\bi_{i} \,.
\label{eq:hp1}
\end{align}
Substituting the transformation in Eq. (\ref{eq:hp1}) into Eq. (\ref{eq:model}), we obtain the following Hamiltonian that is quadratic in magnon operators:
\begin{equation}
\label{eq:h0n}
H_0=-4NJ(1-\lambda)S^2
\end{equation}
\begin{align}
&\frac{H_{2}}{JS} = \sum_{\substack{\langle i,j \rangle \\ i \in A, j \in B}}\left(\hat{a}_i^{\dagger}\hat{a}_i+\hat{b}_j^{\dagger}\hat{b}_j+\hat{a}_i^{\dagger}\hat{b}_j^{\dagger}+\hat{a}_i\hat{b}_j\right)\nonumber\\
& -\frac{\lambda(1+\delta)}{2}\sum_{i\in A} \left(\hat{a}_i^{\dagger}\hat{a}_{i}+\hat{a}_{i\pm\delta_{xy}}^{\dagger}\hat{a}_{i\pm\delta_{xy}}-\hat{a}_i^{\dagger}\hat{a}_{i\pm\delta_{xy}}-\hat{a}_{i}\hat{a}_{i\pm\delta_{xy}}^{\dagger}\right)\nonumber\\
& -\frac{\lambda (1-\delta)}{2}\sum_{i\in A} \left(\hat{a}_i^{\dagger}\hat{a}_{i}+\hat{a}_{i\pm\delta'_{xy}}^{\dagger}\hat{a}_{i\pm\delta'_{xy}}-\hat{a}_i^{\dagger}\hat{a}_{i\pm\delta'_{xy}}-\hat{a}_{i}\hat{a}_{i\pm\delta'_{xy}}^{\dagger}\right)\nonumber\\
& -\frac{\lambda(1+\delta)}{2}\sum_{i\in B} \left(\hat{b}_i^{\dagger}\hat{b}_{i}+\hat{b}_{i\pm\delta'_{xy}}^{\dagger}\hat{b}_{i\pm\delta'_{xy}}-\hat{b}_i^{\dagger}\hat{b}_{i\pm\delta'_{xy}}-\hat{b}_{i}\hat{b}_{i\pm\delta'_{xy}}^{\dagger}\right)\nonumber\\
& -\frac{\lambda (1-\delta)}{2}\sum_{i\in B} \left(\hat{b}_i^{\dagger}\hat{b}_{i}+\hat{b}_{i\pm\delta_{xy}}^{\dagger}\hat{b}_{i\pm\delta_{xy}}-\hat{b}_i^{\dagger}\hat{b}_{i\pm\delta_{xy}}-\hat{b}_{i}\hat{b}_{i\pm\delta_{xy}}^{\dagger}\right) \,,
\label{eq:h2n}
\end{align}
where, $\delta_{xy} \equiv \hat{x}+\hat{y}$ and $\delta'_{xy} \equiv \hat{x}-\hat{y}$ are the directions along the second neighbors.
Using the translation invariance we perform Fourier transform and obtain the following Hamiltonian in the momentum space:
\begin{equation}
\frac{H_{2k}}{S}= \sum_{\kk} \bigg[ J^{aa} \ad_{\kk} \ai_{\kk} + J^{bb} \bd_{\kk} \bi_{\kk} + J^{ab} \left( \ad_{\kk} \bd_{\kk} + \ai_{\kk} \bi_{\kk} \right) \bigg] \,,
\label{eq:h2nk}
\end{equation}
where, 
\begin{align}
\label{eq:jaa}
J^{aa} &= 4J(1+\lambda (\gamma_{\kk}^a-1)+\lambda\delta\gamma_{\kk}^b) \,, \\ 
\label{eq:jbb}
J^{bb} &= 4J(1+\lambda (\gamma_{\kk}^a-1)-\lambda\delta\gamma_{\kk}^b) \,, \\ 
\label{eq:jab}
J^{ab} &= 4J\gamma_{\kk} \,.
\end{align}
In the above expressions, $\gk = (\cos k_{x} + \cos k_{y})/2$, $\gka = (\cos (k_{x} + k_{y}) + \cos(k_{x}-k_{y}))/2$ and $\gkb = (\cos (k_{x} + k_{y}) - \cos(k_{x}-k_{y}))/2$. 
The Hamiltonian in Eq. (\ref{eq:h2nk}) can be straightforwardly diagonalized using Bosonic Bogoliubov transformation, 
\begin{equation}
\ai_{\kk} = u_{\kk} \alpha_{\kk} + v_{\kk} \beta_{\kk}^{\dagger} \,, ~~~
\bi_{\kk} = u_{\kk} \beta_{\kk} + v_{\kk} \alpha_{\kk}^{\dagger}
\label{eq:btn}
\end{equation}
where,
\begin{align}
u_{\kk}^{2} &= \frac{1}{2}\left[\frac{J^{aa}+J^{bb}}{\sqrt{(J^{aa}+J^{bb})^2-4(J^{ab})^2}}+1\right] \,, \nonumber \\
v_{\kk}^{2} &= \frac{1}{2}\left[\frac{J^{aa}+J^{bb}}{\sqrt{(J^{aa}+J^{bb})^2-4(J^{ab})^2}}-1\right] \,, \nonumber \\
u_{\kk} v_{\kk} &= -\frac{J^{ab}}{\sqrt{(J^{aa}+J^{bb})^2-4(J^{ab})^2}} \,.
\label{eq:uvn}
\end{align}
Note that $u_{\kk}^{2} - v_{\kk}^{2} =1$, which makes this transformation anti-unitary. Using the transformation in Eq. (\ref{eq:btn}) we obtain, 
\begin{align}
H_{2k} &= \sum_{\kk} \bigg[ \omega_{1 \kk} ~\alpha_{\kk}^{\dagger} \alpha_{\kk}
+ \omega_{2 \kk} ~\beta_{\kk}^{\dagger} \beta_{\kk} \bigg] \nonumber\\
& +\frac{S}{2} \sum_{\kk}\left[-(J^{aa}+J^{bb})+\sqrt{(J^{aa}+J^{bb})^2-4(J^{ab})^2}\right]  \,,
\label{eq:h2nd}
\end{align}
where the magnon dispersions are 
\begin{align} 
\label{eq:w1n}
\omega_{1 \kk} &= \frac{S}{2}\left[ J^{aa}-J^{bb}+\sqrt{(J^{aa}+J^{bb})^2-4(J^{ab})^2}\right] \,, \\
\label{eq:w2n}
\omega_{2 \kk} &= \frac{S}{2}\left[ J^{bb}-J^{aa}+\sqrt{(J^{aa}+J^{bb})^2-4(J^{ab})^2}\right] \,,
\end{align}
consistent with Refs. \cite{alter_nlswt, Consoli_2025}.
From Eqs. (\ref{eq:w1n})-(\ref{eq:w2n}), we see that generically for any finite value of $\delta$ and $\lambda$ there are two distinct magnon modes. At $\delta =0$ or $\lambda=0$ the two magnon modes are degenerate. In Fig. (\ref{fig:wn}), we have plotted the magnon dispersions.  As expected from spontaneous breaking of the continuous spin-rotation symmetry, there are gapless Goldstone modes. 
At $\kk=(0,0)$ and $\kk = (\pi, \pi)$, $\gk= \pm 1 \,, \gka =1$ and $\gkb=0$ leading to $J^{aa} = J^{bb} = |J^{ab}|$, and thus resulting in gapless modes (see Eqs. (\ref{eq:jaa}) - (\ref{eq:jab}) and Eqs. (\ref{eq:w1n}) - (\ref{eq:w2n})). The magnon band splitting is given by
\begin{equation}
\delta \omega_{\kk} \equiv \omega_{2 \kk} - \omega_{1\kk} = S (J^{bb} - J^{aa}) 
= -8JS \lambda \delta \gkb \,,
\label{eq:mags}
\end{equation}
which is non-zero, except along the $k_{x}$ and $k_{y}$ axes. In Fig. (\ref{fig:mags})(a) we have plotted the magnon splitting in the Brillouin zone. 
Also, the splitting increases with increasing $\lambda$ or $\delta$. 
However, the spin-wave velocity is independent of $\delta$ as well as the direction in the Brillouin zone,
\begin{equation}
c = 2JS\sqrt{2-4\lambda} \,.
\label{eq:swcn}
\end{equation}
It is plotted in Fig. (\ref{fig:mags} (b)). However, this is likely a LSWT result only, and non-linear corrections are expected to render corrections to it. Note that the magnon velocity vanishes at the classical transition point, $\lambda=1/2$.

Using Eqs. (\ref{eq:h0n}), (\ref{eq:h2nd})-(\ref{eq:w2n}), we obtain the leading order expression for the ground-state energy,
\begin{align}
\label{eq:E0n}
E_0 &=
\frac{S}{2} \sum_{\kk}\left[-(J^{aa}+J^{bb})+\sqrt{(J^{aa}+J^{bb})^2-4(J^{ab})^2}\right] \nonumber \\
&-4NJ(1-\lambda)S^2  \,,
\end{align}
which is again independent of $\delta$.

\subsection{Order parameter}
\label{sec:op_af}

\begin{figure}
\centering
\includegraphics[width=0.5\linewidth]{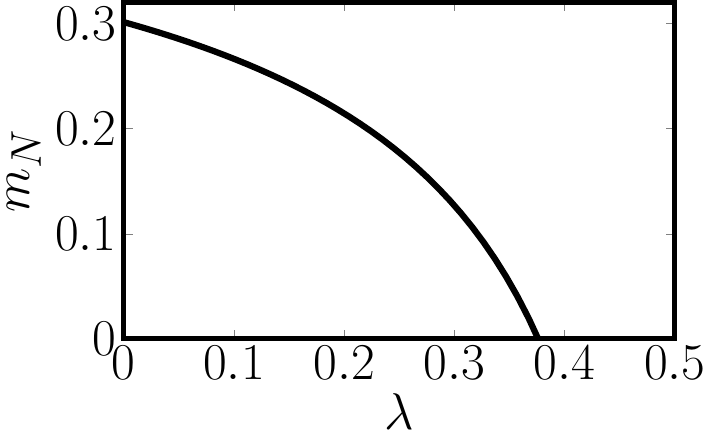}
\caption{Altermagnet/N\'{e}el order parameter (Eqs. (\ref{eq:opN}) and (\ref{eq:opN1})) as a function of $\lambda$ within linear spin-wave theory. As shown in the text, at the level of LSWT quantum corrections to the N\'{e}el order parameter do not have any $\delta$ dependence.  
}
\label{fig:op_af}
\end{figure}

The AM phase is characterized by staggered magnetization,
\begin{equation}
\label{eq:opN}
m_{N} = \frac{1}{N} \sum_{i} \frac{\braket{S^{z}_{Ai}} - \braket{S^{z}_{Bi}}}{2} \,.
\end{equation}
Classically, $m_{N} = S$. Below we calculate the leading corrections to order parameter, i.e., include Gaussian fluctuations,
\begin{align}
\label{eq:opN1}
m_{N} &= S - \Delta m_{N}  
= S - \int_{BZ}\frac{d \kk}{(2\pi)^2} v_{\kk}^{2} \nonumber \\
&= S - \frac{1}{2}\int_{BZ}\frac{d \kk}{(2\pi)^2}
\left[\frac{S(J^{aa}+J^{bb})}{\omega_{1\kk} + \omega_{2\kk}}-1\right] \,,
\end{align}
where the momentum integral is over the full lattice Brillouin zone. 
The source of these quantum corrections are magnon excitations. As we have seen earlier, for finite value of $\delta$ there are two distinct magnon modes and the dispersion relation depends on $\delta$. However, we find that the quantum correction, $\Delta m_{N}$,  is independent of $\delta$ at the LSWT order. Thus in the view of quantum corrections at this level, the magnon splitting in altermagnet has no additional consequence compared to the usual N\'{e}el antiferromagnet phase realized at $\delta=0$. 
This fact was also noted recently in Ref. \cite{alter_nlswt}. In Fig. (\ref{fig:op_af}) we show the variation of $m_{N}$ as a function of $\lambda$. We find that the AM phase becomes unstable at $\lambda \approx 0.375$, independent of $\delta$. 
This is probably an effect only at the linear spin-wave theory level, and most likely the quantum corrections will depend on $\delta$ upon including higher-order corrections. 

\subsection{Dynamic spin structure factor}
\label{sec:dsf_af}

\begin{figure}[t]
\centering
\includegraphics[width=0.48\linewidth]{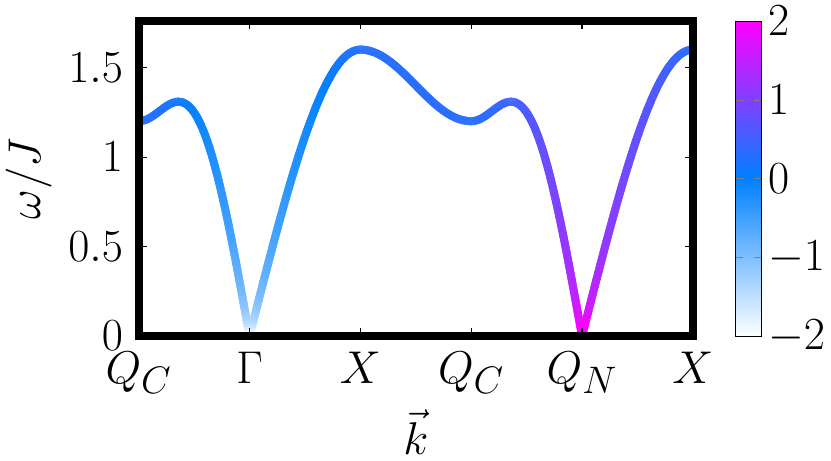}
\includegraphics[width=0.48\linewidth]{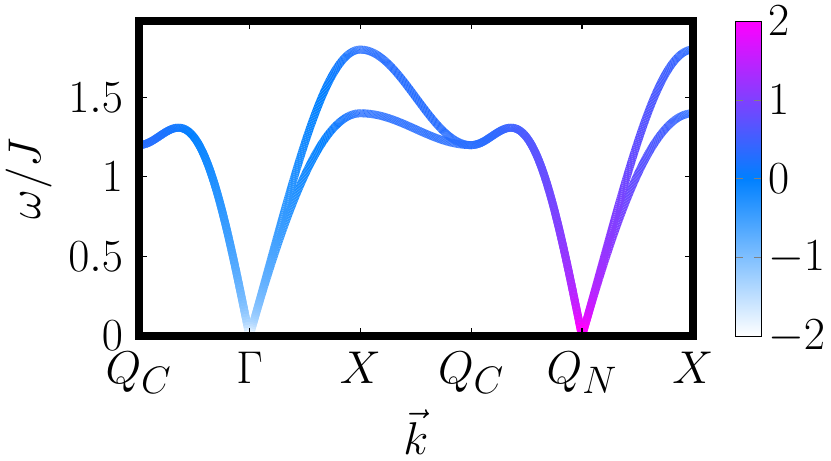}
\caption{Dynamic spin structure factor (Eq. (\ref{eq:dsf_af})) in AM phase is plotted for different parameters. (a) $\delta=0$,$\lambda=0.2$; (b) $\delta=0.5$,$\lambda=0.2$. At $\delta=0$ the magnon modes are degenerate while they split for $\delta \neq 0$. Moreover, the spectral weight is peaked at $\vec{Q}_{N} = (\pi,\pi)$, as expected for AM order on the square lattice.}
\label{fig:dsf}
\end{figure}

\begin{figure}[t]
\centering
\includegraphics[width=0.7\linewidth]{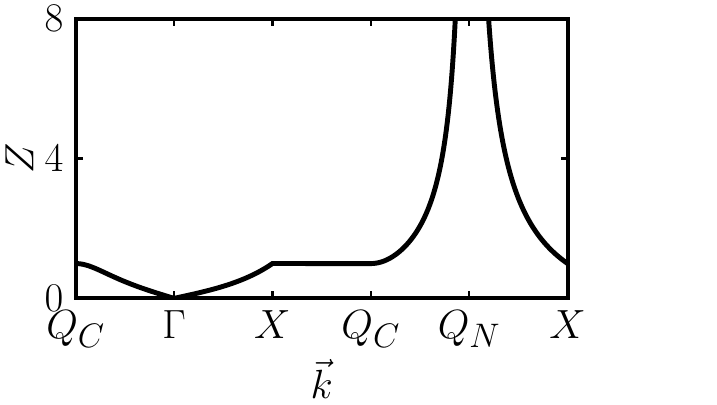}
\caption{Spectral weight in the dynamic structure factor in the AM phase at $\lambda=0.2$, $\delta=0.5$.
}
\label{fig:ssf}
\end{figure}

In this section we discuss the dynamic spin structure factor, which is accessible in inelastic neutron scattering experiment. The 
dynamic spin structure factor is given by
\begin{equation} 
\label{eq:dsf_def}
\mathcal{S}(\kk,\omega) 
= \frac{1}{2\pi N_{s}} \sum_{ij} \int_{-\infty}^{\infty}dt e^{i(\omega t - \kk\cdot(\vec{r}_{i}-\vec{r}_{j}) )} 
\langle \vec{S}_{i} (t) \cdot \vec{S}_{j} (0) \rangle \,,
\end{equation}
where $N_{s} = 2N$ is the total number of spins.
In the AM phase, at $T=0$, we obtain,
\begin{align}
\label{eq:dsf_af}
\mathcal{S}(\kk,\omega) &= Z_{1\kk} ~\delta(\omega-\omega_{1\kk}) + Z_{2\kk} ~\delta(\omega-\omega_{2\kk})  \,, \\ 
Z_{1\kk} = Z_{2\kk} &= 2S\sqrt{\frac{J^{aa}+J^{bb}-2J^{ab}}{J^{aa}+J^{bb}+2J^{ab}}} \,.
\end{align}
From the above expressions we see that the spectral weight is independent of $\delta$ (see Eqs. (\ref{eq:jaa})--(\ref{eq:jab})) and it is same for both the magnon modes. This is again most likely an effect only at the LSWT level and the spectral weight is expected to vary with $\delta$ when magnon-magnon interactions are taken into account. In Fig. (\ref{fig:dsf}) we plot the dynamic spin structure factor, while in Fig. (\ref{fig:ssf}) we plot the spectral weight. As expected, the maximum spectral weight is at the ordering wavevector, $\vec{Q}_{N} = (\pi,\pi)$, while it vanishes at the $\Gamma \equiv (0,0)$ point, which is the other gapless point.

The local spin susceptibility is obtained from the dynamic spin structure factor as follows:
\begin{equation}
\mathcal{S}(\omega)=\int_{BZ} d\kk ~\mathcal{S}(\kk,\omega) \,.
\end{equation}
In general, this needs to be calculated numerically but at $\omega=0$ the contributing points are $\kk= (0,0)$ and $\kk= (\pi,\pi)$ making it straightforward to evaluate analytically. Thus, 
\begin{equation}
\mathcal{S}(0)=\frac{4\pi}{J(1-2\lambda)} \,.
\end{equation}
For low frequencies, the local spin susceptibility is independent of $\omega$, as expected for a magnetically ordered phase. 
Note that the local spin susceptibility is again $\delta$ independent at $\omega=0$. 
We can also evaluate the static structure factor by momentum integration of the dynamic structure factor. 


\section{Columnar phase}
\label{sec:col}

\begin{figure}
\centering
\subfloat[]{\includegraphics[width=0.5\linewidth]{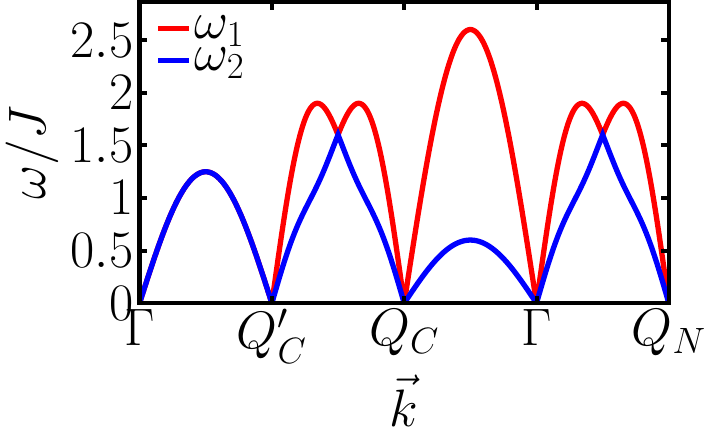}}
\subfloat[]{\includegraphics[width=0.5\linewidth]{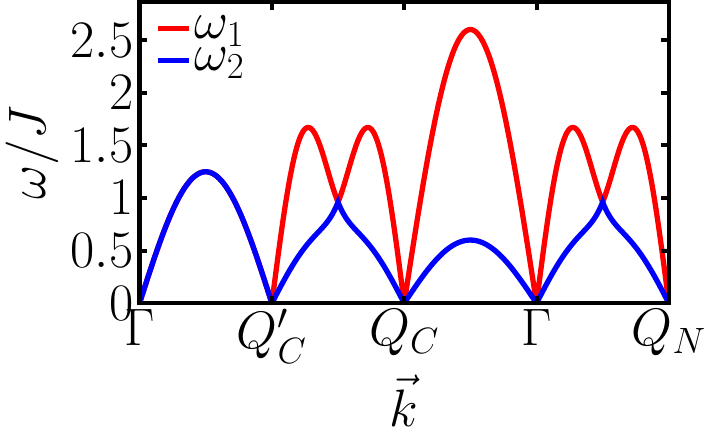}} \\
\subfloat[]{\includegraphics[width=0.5\linewidth]{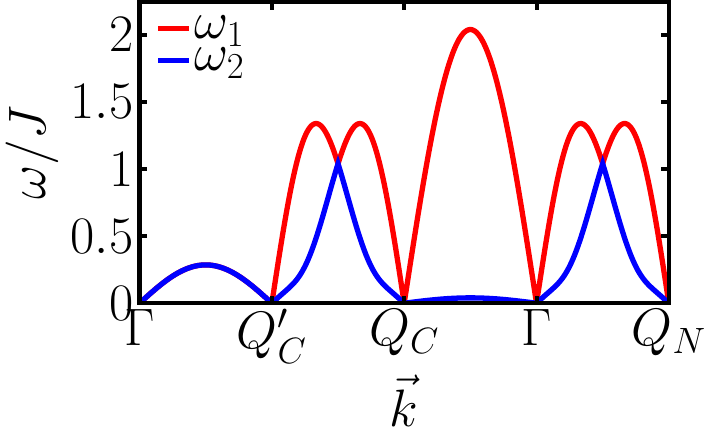}}
\subfloat[]{\includegraphics[width=0.5\linewidth]{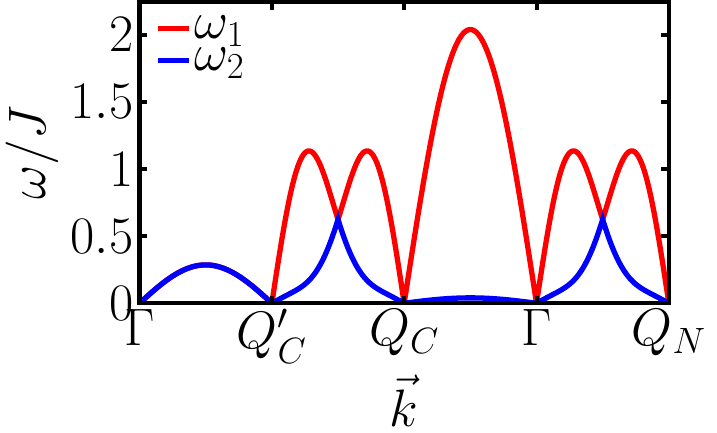}}
\caption{CAF phase dispersion. Unlike the Altermagnet case there is no magnon band splitting here. The two magnon modes seen here are simply the consequence of larger magnetic unit cell. (a) $\delta=0, \lambda=0.8$, (b) $\delta=0.8, \lambda=0.8$, (c) $\delta=0, \lambda=0.52$ (d) $\delta=0.8, \lambda=0.52$. 
}
\label{fig:wc}
\end{figure}

We now turn our attention to the CAF phase realized for $\lambda > 1/2$. As shown in Fig. (\ref{fig:model} (c)), the magnetic unit cell now contains $4$ spins. The Holstein-Primakoff transformation on these $4$ sublattices is given as follows: 
\begin{align}
\label{eq:hpc}
S^{z}_{Ai} &= S - \ad_{i} \ai_{i} \,, && S^{z}_{Bi} = \bd_{i} \bi_{i} - S \,, \nonumber \\ 
S^{+}_{Ai} &= \sqrt{2S - \ad_{i}\ai_{i}} ~\ai_{i} \,, && S^{+}_{Bi} = \bd_{i} \sqrt{2S - \bd_{i}\bi_{i}}  \,, \nonumber \\ 
S^{-}_{Ai} &= \ad_{i} \sqrt{2S - \ad_{i}\ai_{i}} \,, && S^{-}_{Bi} = \sqrt{2S - \bd_{i}\bi_{i}} ~\bi_{i} \,, \nonumber \\
S^{z}_{Ci} &= \cd_{i} \ci_{i} - S \,, && S^{z}_{Di} = S - \dd_{i} \di_{i}  \,, \nonumber \\ 
S^{+}_{Ci} &= \cd_{i} \sqrt{2S - \cd_{i}\ci_{i}}  \,, && S^{+}_{Di} = \sqrt{2S - \dd_{i}\di_{i}} ~\di_{i} \,,  \nonumber \\ 
S^{-}_{Ci} &= \sqrt{2S - \cd_{i}\ci_{i}} ~\ci_{i} \,, && S^{-}_{Di} = \dd_{i} \sqrt{2S - \dd_{i}\di_{i}} \,,
\end{align}
where $i$ is the unit cell index. 
As discussed in Sec. \ref{sec:neel}, we work within the LSWT in which the above transformation simplifies to the following:
\begin{align}
\label{eq:hpc1}
S^{z}_{Ai} &= S - \ad_{i} \ai_{i} \,, && S^{z}_{Bi} = \bd_{i} \bi_{i} - S \,, \nonumber \\ 
S^{+}_{Ai} &\approx \sqrt{2S} ~\ai_{i} \,, && S^{+}_{Bi} \approx \sqrt{2S} ~\bd_{i}  \,, \nonumber \\ 
S^{-}_{Ai} &\approx \sqrt{2S} ~\ad_{i} \,, && S^{-}_{Bi} \approx \sqrt{2S} ~\bi_{i} \,, \nonumber \\
S^{z}_{Ci} &= \cd_{i} \ci_{i} - S \,, && S^{z}_{Di} = S - \dd_{i} \di_{i} \,, \nonumber \\ 
S^{+}_{Ci} &\approx \sqrt{2S} ~\cd_{i}  \,, && S^{+}_{Di} \approx \sqrt{2S} ~\di_{i} \,,\nonumber \\ 
S^{-}_{Ci} &\approx \sqrt{2S} ~\ci_{i} \,, && S^{-}_{Di} \approx \sqrt{2S} ~\dd_{i} \,.
\end{align}
Following the same steps as in the earlier section we obtain the Hamiltonian in terms of magnon operators at the quadratic level,
\begin{align}
\label{eq:h0c}
&H_0=-8NJ\lambda S^2 \\
\label{eq:h2c}
&\frac{H_{2}}{JS} = \sum_{\substack{\langle i,j \rangle \\ i \in A, j \in B}}\left(\hat{a}_i^{\dagger}\hat{a}_i+\hat{b}_j^{\dagger}\hat{b}_j+\hat{a}_i^{\dagger}\hat{b}_j^{\dagger}+\hat{a}_i\hat{b}_j\right)\nonumber\\
& -\sum_{\substack{\langle i,j \rangle \\ i \in A, j \in D}}\left(\hat{a}_i^{\dagger}\hat{a}_i+\hat{d}_j^{\dagger}\hat{d}_j-\hat{a}_i\hat{d}_j^{\dagger}-\hat{a}_i^{\dagger}\hat{d}_j\right)\nonumber\\
& + \sum_{\substack{\langle i,j \rangle \\ i \in C, j \in D}}\left(\hat{c}_i^{\dagger}\hat{c}_i+\hat{d}_j^{\dagger}\hat{d}_j+\hat{c}_i^{\dagger}\hat{d}_j^{\dagger}+\hat{c}_i\hat{d}_j\right)\nonumber\\
& -\sum_{\substack{\langle i,j \rangle \\ i \in C, j \in B}}\left(\hat{c}_i^{\dagger}\hat{c}_i+\hat{b}_j^{\dagger}\hat{b}_j-\hat{c}_i\hat{b}_j^{\dagger}-\hat{c}_i^{\dagger}\hat{b}_j\right)\nonumber\\
& +\lambda(1+\delta)\sum_{i \in A} \left(\hat{a}_i^{\dagger}\hat{a}_i+\hat{c}_{i\pm\delta_{xy}}^{\dagger}\hat{c}_{i\pm\delta_{xy}}+\hat{a}_i^{\dagger}\hat{c}_{i\pm\delta_{xy}}^{\dagger}+\hat{a}_i\hat{c}_{i\pm\delta_{xy}}\right)\nonumber\\
& +\lambda(1-\delta)\sum_{i \in A} \left(\hat{a}_i^{\dagger}\hat{a}_i+\hat{c}_{i\pm\delta'_{xy}}^{\dagger}\hat{c}_{i\pm\delta'_{xy}}+\hat{a}_i^{\dagger}\hat{c}_{i\pm\delta'_{xy}}^{\dagger}+\hat{a}_i\hat{c}_{i\pm\delta'_{xy}}\right)\nonumber\\
& +\lambda(1+\delta)\sum_{i \in B} \left(\hat{b}_i^{\dagger}\hat{b}_i+\hat{d}_{i\pm\delta'_{xy}}^{\dagger}\hat{d}_{i\pm\delta'_{xy}}+\hat{b}_i^{\dagger}\hat{d}_{i\pm\delta'_{xy}}^{\dagger}+\hat{b}_i\hat{d}_{i\pm\delta'_{xy}}\right)\nonumber\\
& +\lambda(1-\delta)\sum_{i \in B} \left(\hat{b}_i^{\dagger}\hat{b}_i+\hat{d}_{i\pm\delta_{xy}}^{\dagger}\hat{d}_{i\pm\delta_{xy}}+\hat{b}_i^{\dagger}\hat{d}_{i\pm\delta_{xy}}^{\dagger}+\hat{b}_i\hat{d}_{i\pm\delta_{xy}}\right) \,.
\end{align}
After performing a Fourier transformation, we obtain,
\begin{align} 
\frac{H_{2k}}{JS} &= \sum_{\kk}  4\lambda \left( {\ad_{\kk}} \ai_{\kk} +   {\bd_{\kk}} \bi_{\kk} 
+ {\cd_{\kk}} \ci_{\kk} + {\dd_{\kk}} \di_{\kk} \right)  \nonumber \\ 
&+ \sum_{\kk} \Bigg[ 
2\cos{k_{x}} \left( \ad_{\kk} \di_{\kk} + \cd_{\kk} \bi_{\kk} \right)
+ 2\cos{k_{y}} \left( \ad_{\kk} \bd_{\kk}  + \cd_{\kk} \dd_{\kk} \right)  \nonumber\\ 
&+ 4\lambda \big[\gamma_{\kk}^a-\delta\gamma_{\kk}^b \big] \ad_{\kk} \cd_{\kk} +4\lambda \big[\gamma_{\kk}^a+\delta\gamma_{\kk}^b \big] \bd_{\kk} \dd_{\kk} + \text{H.c.} \Bigg] \,.
\label{eq:h2ck} 
\end{align}
We first express the above Hamiltonian in the Bogoliubov-de Gennes form,
\begin{equation}
\label{eq:bdgc}
\frac{H_{2k}}{JS} = \sum_{\kk} \psi_{\kk}^{\dagger} h_{\kk} \psi_{k} - 8\lambda N    \,,
\end{equation}
where $\psi_{\kk}^{\dagger} = \begin{pmatrix}
    \bd_{\kk} & \cd_{\kk} & \ai_{\kk} & \di_{\kk}
\end{pmatrix} $ 
and 
\begin{equation}
\label{eq:hkc}
h_{\kk} = 
\begin{bmatrix}
h_{1\kk} && h_{2\kk} \\
h_{2\kk}^{\dagger} && h_{1\kk}
\end{bmatrix} \,,
\end{equation}
with
$h_{1 \kk} = 4 \lambda \mathbbm{1} + 2\cos{k_{x}} \sigma_{1}$ and $h_{2\kk} = 2 \cos{k_{y}} \mathbbm{1} + 4\lambda \gka \sigma_{1} + i 4\lambda \delta \gkb \sigma_{2}$.
As in the case of the altermagnetic phase, the above Hamiltonian can be straightforwardly diagonalized using Bosonic Bogoliubov transformation, the details of which are presented in Appendix \ref{sec:appC}. 
The magnon mode dispersion is given by the eigenvalues of the non-hermitian matrix $\Sigma h_{\kk}$, where $\Sigma = diag(1,1,-1,-1)$, which follows from the anti-unitary Bogoliubov transformation. The evaluation of these eigenvalues is straightforward and we obtain the magnon modes as follows:
\begin{align}
\label{eq:w1c}
&\frac{\omega_{1 \kk}}{2JS}= \Bigg[4\lambda^2 + \left(\cos^2{k_x}-\cos^2{k_y}\right) -4\lambda^2\left((\gka)^2+\delta^2(\gkb)^2\right)\nonumber \\ &+4\lambda \sqrt{4\lambda^2\delta^2(\gka)^2(\gkb)^2-\delta^2(\gkb)^2\cos^2{k_x}+\cos^2{k_x}\sin^4{k_y}}\Bigg]^\frac{1}{2}\\
\label{eq:w2c}
&\frac{\omega_{2 \kk}}{2JS}= \Bigg[4\lambda^2 + \left(\cos^2{k_x}-\cos^2{k_y}\right) -4\lambda^2\left((\gka)^2+\delta^2(\gkb)^2\right)\nonumber \\ &-4\lambda \sqrt{4\lambda^2\delta^2(\gka)^2(\gkb)^2-\delta^2(\gkb)^2\cos^2{k_x}+\cos^2{k_x}\sin^4{k_y}}\Bigg]^\frac{1}{2} \,,
\end{align}
From Eqs. (\ref{eq:w1c})-(\ref{eq:w2c}), we see that generically for any value of $\delta$ and $\lambda$ there are two distinct doubly degenerate magnon modes owing to the different sublattices in the magnetic unit cell. 
In Fig. (\ref{fig:wc}), we have plotted the magnon dispersions. As expected from spontaneous breaking of spin-rotation symmetry there is a gapless Goldstone mode at the ordering wavevector ($\vec{Q}_{c} = (0,\pi)$). In addition there are gapless modes at the $\Gamma$ point, $\vec{Q}_{N} = (\pi,\pi)$ and $\vec{Q}_{c}'=(\pi,0)$.

\begin{figure}[t]
\centering
\subfloat[]{\includegraphics[width=0.49\linewidth]{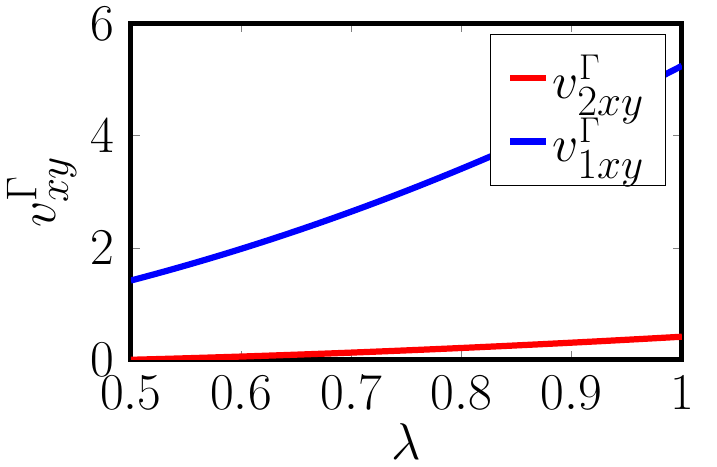}}
\subfloat[]{\includegraphics[width=0.49\linewidth]{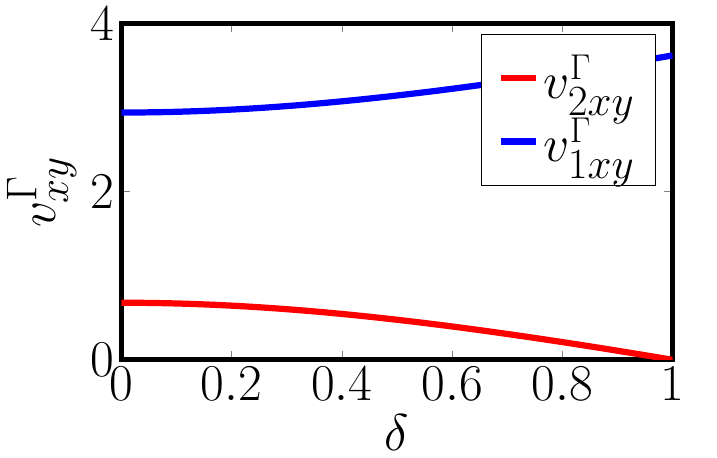}}
\caption{Magnon velocities in the CAF phase around the $\Gamma$ point. (a) $\delta=0.8$, (b) $\lambda=0.8$.
}
\label{fig:swc1}
\end{figure}

\begin{figure}
\centering
\subfloat[]{\includegraphics[width=0.49\linewidth]{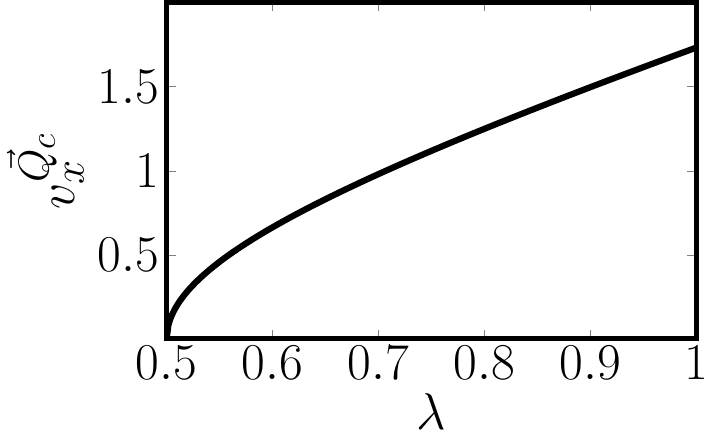}}
\subfloat[]{\includegraphics[width=0.49\linewidth]{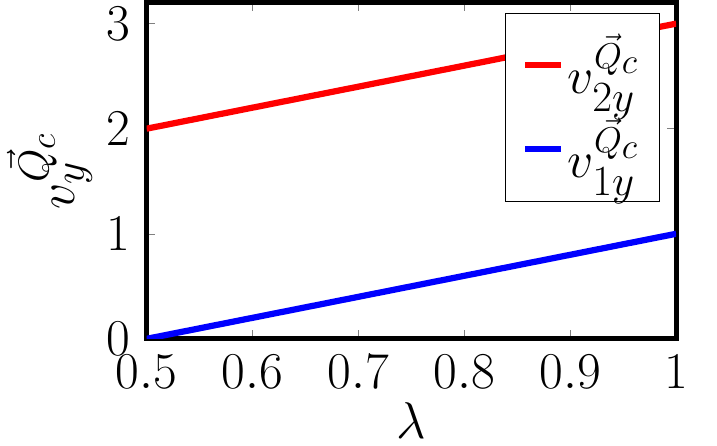}}
\caption{Magnon velocities in the CAF phase around the $\vec{Q}_{c}$. (a) Magnon velocity along $k_{x}$. (b) Magnon velocity along $k_{y}$. Both the quantities are independent of $\delta$, as discussed in the main text. 
}
\label{fig:swc2}
\end{figure}

Unlike the AM case, here the magnon velocities are different for the two modes and importantly in general depend on the direction in the BZ as well as $\delta$. At the classical transition point, $\lambda = 1/2$, the magnon velocity of the lower magnon band along the $\vec{Q}_{c} -\Gamma$ and $\vec{Q}'_{c} -\Gamma$ paths vanishes. The magnon velocity of the lower band also vanishes at $\delta=1$. Below are the specific expressions of the magnon velocities and these are also plotted in Figs. (\ref{fig:swc1}) - (\ref{fig:swc2}).
Around $\vec{Q}_{c}$, the magnon velocities of the two bands along $k_{x}$ are identical, whereas they are different along the $k_{y}$ direction,
\begin{equation}
\label{columnspinwavevel1}
v_{1,2 x}^{\vec{Q}_c}=2JS\sqrt{4\lambda^2-1} \,, \quad v_{1,2 y}^{\vec{Q}_c}=2JS(2\lambda\mp 1) \,.
\end{equation}
However, these are independent of $\delta$.
Around the $\Gamma$ point, the magnon velocities are $\delta$ dependent along the $\Gamma - \vec{Q}_{N}$ path,
\begin{align}
\label{columnspinwavevel3}
v_{1,2 xy}^{\Gamma}=2\sqrt{2}JS
\sqrt{2\lambda^2\pm\lambda\sqrt{4\lambda^2\delta^2-\delta^2+1}}
\end{align}
The magnon velocities are shown in Figs. (\ref{fig:swc1}) and (\ref{fig:swc2}).

Finally, using Eqs. (\ref{eq:h0c}), (\ref{eq:h2ck})-(\ref{eq:w2c}), we obtain the ground-state energy at the LSWT level,
\begin{equation}
\label{eq:E0c}
E_0=-8J\lambda NS(1+S)+JS\sum_{\kk}(\omega_{1\kk}+\omega_{2\kk}) \,.
\end{equation}

\subsection{Order parameter}
\label{sec:op_caf}

\begin{figure}
\centering
\subfloat[]{\includegraphics[width=0.5\linewidth]{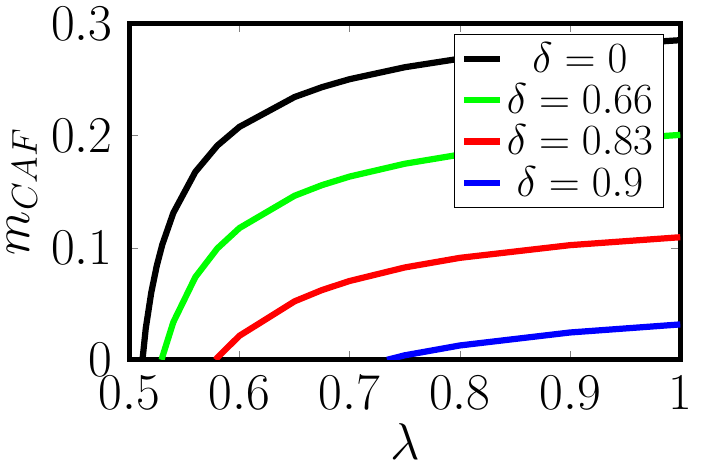}} 
\subfloat[]{\includegraphics[width=0.5\linewidth]{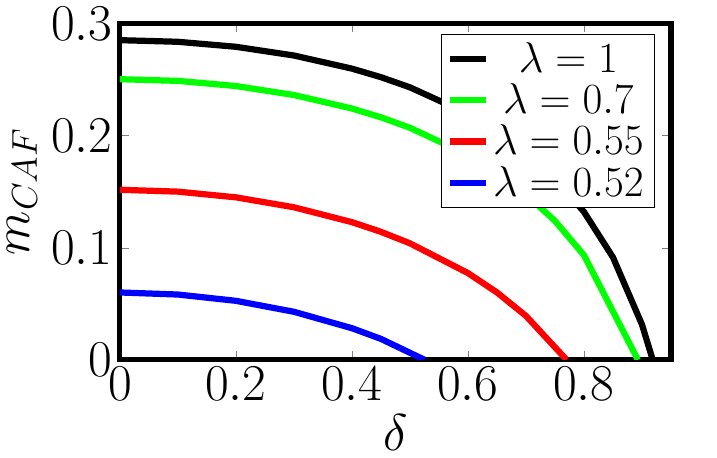}
}
\caption{Order parameter $m_{CAF}$ corresponding to the CAF phase (see Eqs. (\ref{eq:opC}) -- \ref{eq:opC4}) plotted as a function of $\lambda$. With increasing value of $\delta$ this phase becomes more unstable.
(b) $m_{CAF}$ plotted as a function of $\delta$ for fixed values of $\lambda$.
}
\label{fig:opc}
\end{figure}

The CAF phase is characterized by the order parameter,
\begin{align}
\label{eq:opC}
m_{CAF} &= \frac{1}{N} \sum_{i} \frac{ \braket{S^{z}_{Ai}} - \braket{S^{z}_{Bi}} - \braket{S^{z}_{Ci}} + \braket{S^{z}_{Di}} }{4} \,, \nonumber \\ 
&\equiv \frac{m_{A} - m_{B} - m_{C} + m_{D}}{4} \,.
\end{align}
Below we calculate the leading corrections to order parameter, i.e., include Gaussian fluctuations. In terms of the Bogoliubov coefficients (see Appendix \ref{sec:appC} for details),
\begin{align}
\label{eq:opC1}
m_{A} &= 
S -  \frac{1}{\pi^2}\int_{-\frac{\pi}{2}}^{\frac{\pi}{2}}dk_x\int_{-\frac{\pi}{2}}^{\frac{\pi}{2}}dk_y (t_{31}^2(k)+t_{32}^2(k))  \,, \\
\label{eq:opC2}
m_{B} &=
S -  \frac{1}{\pi^2}\int_{-\frac{\pi}{2}}^{\frac{\pi}{2}}dk_x\int_{-\frac{\pi}{2}}^{\frac{\pi}{2}}dk_y (t_{13}^2(k)+t_{14}^2(k))  \,,\\
\label{eq:opC3}
m_{C} &=
S -  \frac{1}{\pi^2}\int_{-\frac{\pi}{2}}^{\frac{\pi}{2}}dk_x\int_{-\frac{\pi}{2}}^{\frac{\pi}{2}}dk_y (t_{23}^2(k)+t_{24}^2(k))  \,,\\
\label{eq:opC4}
m_{D} &=
S -  \frac{1}{\pi^2}\int_{-\frac{\pi}{2}}^{\frac{\pi}{2}}dk_x\int_{-\frac{\pi}{2}}^{\frac{\pi}{2}}dk_y (t_{41}^2(k)+t_{42}^2(k))  \,,
\end{align}
Unlike the altermagnet phase, here the quantum corrections  depend on $\delta$, although there is no magnon band splitting. In fact, with increasing $\delta$ the quantum corrections become stronger such that the CAF phase becomes less stable. We find that at $\delta=1$ the CAF does not exist at all. This is a consequence of the vanishing spin-wave velocity, as discussed above. Due to quantum fluctuations, for a fixed value of $\delta$, the order parameter vanishes at a value $\lambda_{c} > 1/2$, and the critical value $\lambda_{c}$ increases with $\delta$. Also, for a fixed value of $\lambda$, the critical value $\delta_{c}$ where the order parameter vanishes decreases. In Fig. (\ref{fig:opc}) we show the variation of $m_{CAF}$ as a function of $\lambda$ as well as $\delta$.  At $\lambda=1$ $\delta_{c}=0.92$, while at $\delta=0$, $\lambda_{c} = 0.51$, which is consistent with the previous studies  \cite{Rojas_2011}. Consequently, there is region of quantum disordered phase where neither the AM phase nor the CAF phase exists. In fact, we find that the domain of the quantum disordered region increases with increasing $\delta$.  

\subsection{Dynamic spin structure factor}
\label{sec:dsf_caf}

\begin{figure}
\centering
\subfloat[]{\includegraphics[width=0.49\linewidth]{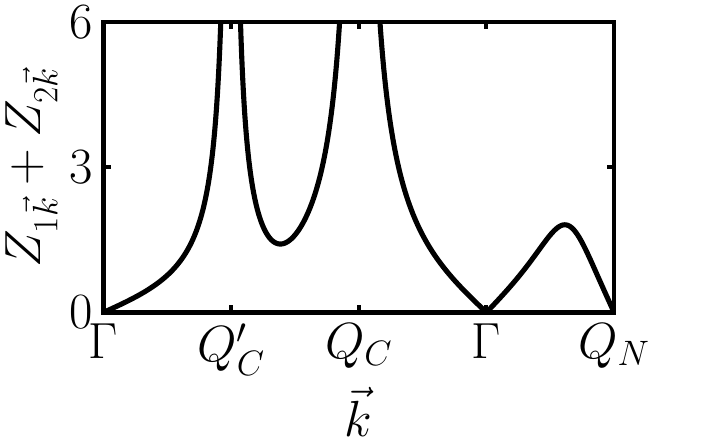}}
\subfloat[]{\includegraphics[width=0.49\linewidth]{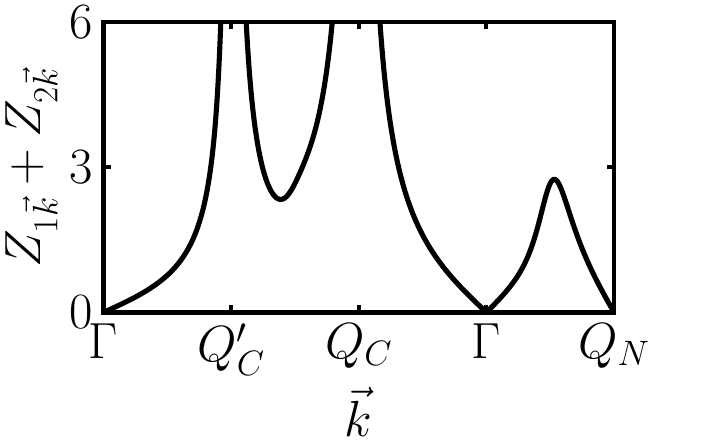}}
\caption{Spectral weights in the CAF phase plotted along a path in the BZ for (a) $\delta=0 \,, \lambda=0.8$ and (b) for $\delta=0.8 \,, \lambda=0.8$.
}
\label{fig:zcaf}
\end{figure}

In this section, we evaluate the dynamic spin structure factor in CAF phase. Using the definition in Eq. (\ref{eq:dsf_def}),
\begin{equation}
\mathcal{S}(\kk,\omega)=
Z_{1\kk} ~\delta(\omega - \omega_{1\kk}) 
+ Z_{2\kk} ~\delta(\omega - \omega_{2\kk}) \,,
\end{equation}
where
\begin{align}
\label{eq:z1c}
Z_{1\kk} =4 \left( \sum_{i=1}^{4} t_{i1}(\kk) \right)^{2} 
+4 \left( \sum_{i=1}^{4} t_{i3}(\kk) \right)^{2} \,, \\
\label{eq:z2c}
Z_{2\kk} =4 \left( \sum_{i=1}^{4} t_{i2}(\kk) \right)^{2} 
+4 \left( \sum_{i=1}^{4} t_{i4}(\kk) \right)^{2} \,,
\end{align}
where $t's$ are the Bogoliubov coefficients, whose details can be found in Appendix \ref{sec:appC}.
In Fig. (\ref{fig:dsf_c}), we plot the dynamic spin structure factor. As expected, the spectral weight is maximum at the ordering wavevector, $\vec{Q}_{C} = (0, \pi)$ for CAF order, whereas it vanishes at the $\Gamma$ and $\vec{Q}_{N}$ points where also the mode is gapless. The spectral weight is also non-zero at $\vec{Q}_{c}' = (\pi,0)$, which is symmetry related to $\vec{Q}_{c}$ and corresponds to an ordering where the ferromagnetic alignment of spins is along the $y-$axis instead of $x-$axis, as considered in the present calculation (see Fig. (\ref{fig:model} (c))). At the LSWT level, the spectral weight diverges at both $\vec{Q}_{c}$ and $\vec{Q}'_{c}$ (as shown in Fig. (\ref{fig:zcaf})). However, with non-linear corrections it is expected that the spectral weight will diverge only at $\vec{Q}_{c}$, while there will be a local maximum at $\vec{Q}'_{c}$. In fact, even slight anisotropy in the NN interactions already leads to divergence only at $\vec{Q}_{c}$ even at the LSWT level \cite{Singh95, Sandvik2001, Applegate10}. 

Here a few remarks are in order. Firstly, for $\delta=0$, the magnetic phase is fully described by only $2$ spins per magnetic unit cell. It was our choice to use $4$ sublattices instead and hence we obtained $2$ magnon mode dispersions (see Eqs. (\ref{eq:w1c})-(\ref{eq:w2c})). However, dynamic spin structure factor, which is a physical observable in inelastic neutron scattering experiment, correctly peaks only at one of the magnon energies, as seen in Fig. (\ref{fig:dsf_c}) (a). For $\delta \neq 0$, the magnetic unit cell necessarily has $4$ spins owing to the two different types of second neighbor interactions (see Fig. (\ref{fig:model}) (c)). Therefore two different magnon bands are expected and this is correctly reflected in the dynamic spin structure factor, as shown in Fig. (\ref{fig:dsf_c}) (b). 
In Fig. (\ref{fig:dsf_c}) (b) , i.e. for $\delta \neq 0$, we see that there appears to be magnon band splitting in regions of BZ, akin to an altermagnet. However, this is slightly subtle.
As stated earlier, an additional magnon mode is expected due to larger magnetic unit cell. These magnon modes are generically not expected to be degenerate and so need not imply altermagnetism. As we approach the limit $\delta \to 0$, the splitting of the magnon bands does not vanish, unlike the AM phase in the previous section. Instead, one of the magnon branches looses the spectral weight leading to a single magnon branch at $\delta =0$. However, we note that for $\delta \neq 0$, the opposite spin sublattices are not trivially connected by simple lattice translation and time-reversal, as opposed to the $\delta=0$ case. Therefore, a more careful analysis of magnetic symmetry group and chirality of magnons is required to determine if the magnon splitting for $\delta \neq 0$ is that of altermagnet. 
Note that in the AM phase discussed in the previous section,
this issue does not arise as the magnetic unit cell has $2$ spins for both $\delta=0$ as well as $\delta \neq 0$.

\begin{figure}
\centering
\subfloat[]{\includegraphics[width=0.45\linewidth]{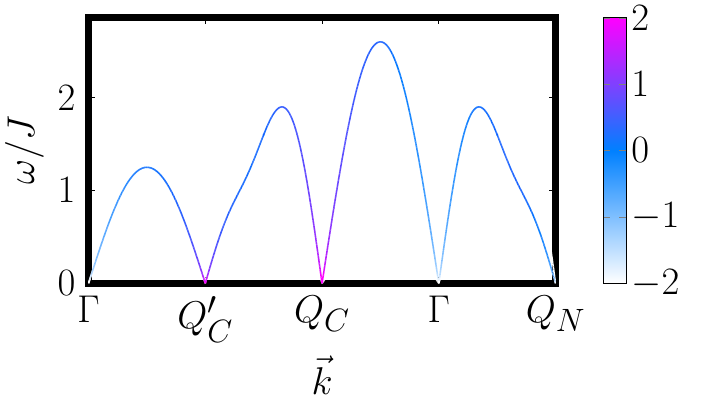}}
\subfloat[]{\includegraphics[width=0.45\linewidth]{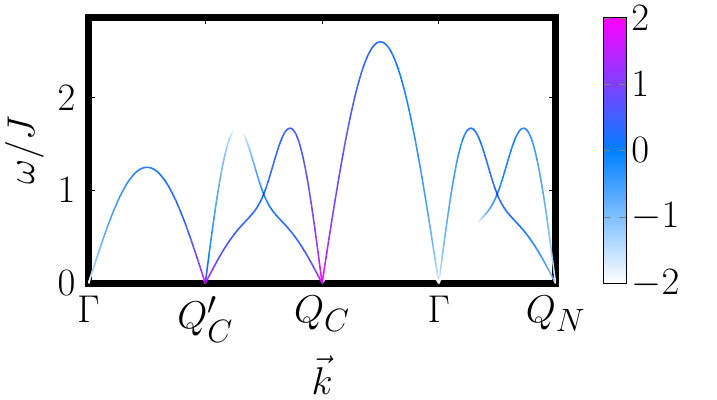}}
\caption{Dynamic spin structure factor in the CAF phase.
(a) $\delta=0 \,, \lambda=0.8$. Here the magnetic state has only one physically observable mode. (b) $\delta=0.8 \,, \lambda=0.8$. Here the magnetic state has $4$ sublattices and correspondingly $2$ magnon modes are visible.
} 
\label{fig:dsf_c}
\end{figure}

\begin{figure}[t]
\centering
\includegraphics[width=0.6\linewidth]{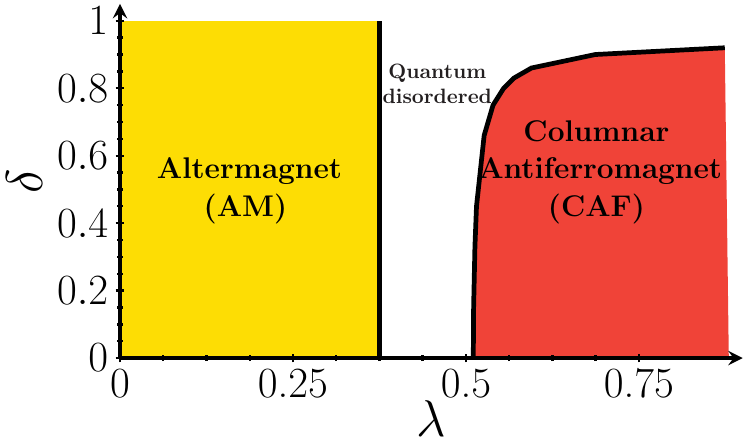}
\caption{Phase diagram corresponding to the model considered in Eq. (\ref{eq:model}), obtained using LSWT. The AM is stable for $\lambda \lesssim 0.375$, while the CAF is stable for $\lambda \gtrsim 0.511$. There is an intermediate quantum disordered phase whose stability increases with increasing $\delta$. 
}
\label{fig:pd}
\end{figure}

\section{Summary}
\label{sec:con}

We have studied a Heisenberg model on a square lattice with two different types of second-neighbor interactions (see Eq. (\ref{eq:model})). This provides us another tuning parameter for frustration. As discussed earlier, the model smoothly interpolates between two well studied models, namely the $J_{1} - J_{2}$ model on square lattice and the checkerboard lattice model. Using LSWT, we have calculated magnon modes, order parameter as well as dynamic spin structure factor. Combining the results obtained in Secs. \ref{sec:neel} and \ref{sec:col}, we find that the magnetically ordered phases become unstable due to quantum fluctuations, and there is a wide region of quantum disordered phase that emerges. The altermagnet phase realized for small $\lambda < 1/2$ is unaffected by $\delta$ anisotropy in our calculation. This is expected as a leading order result because the weakening of second neighbor bonds on alternate plaquette doesn't lead to additional frustration for N\'{e}el type spin order. However, we expect that magnon-magnon interactions will change this situation. What is interesting to note here is that at least at this level of leading order results, in terms of quantum fluctuations, there is no difference between the standard N\'{e}el antiferromagnet and an altermagnet. This is also clear from the fact that the spin-wave velocities of the split magnon bands is identical. So the non-trivial difference between the standard N\'{e}el and an altermagnet is truly a non-linear effect, which should be seen upon introducing magnon-magnon interaction.

The CAF phase, on the other hand, is very sensitive to the frustration caused by $\delta \neq 0$ anisotropy. This is again straightforward to see at this leading order. The CAF phase is stabilized by the presence of the second-neighbor interactions. Therefore, weakening of second neighbor bonds on alternating plaquettes directly impacts the stability of CAF phase.

Based on these results we obtain the phase diagram shown in Fig. (\ref{fig:pd}), where apart from the magnetically ordered phases discussed here there is a wide region of quantum disordered phase. This quantum disordered phase is connected to the one well known in the $J_{1}-J_{2}$ model. In fact, with increasing the new frustration parameter, $\delta$, in our model this quantum disordered region is further stabilized. 
Our method does not allow us to determine the exact nature of this quantum disordered phase. However, given that the $J_{1}-J_{2}$ model hosts dimer-ordered and QSL phases, we expect similar phases to be present for $\delta \neq 0$. Because the second-neighbor interactions on alternating plaquettes are different and with increasing $\delta$ one type of plaquette has stronger bonds than the other, we may expect that the plaquette valence-bond phase to be more stabilized compared to the columnar valence-bond phase. A thorough analysis of this possibility is left for future work. Recently, the quantum disordered phase has been investigated numerically with the possibility of a bond-nematic phase \cite{Duric26}. 

What is interesting would be to investigate the stability of the QSL phase found in the $J_{1}-J_{2}$ model, with increasing $\delta$. Since the lattice symmetry is reduced with $\delta \neq 0$, on general symmetry grounds one may expect that the QSL found at $\delta =0$ may be immediately unstable. But there is an interesting possibility of another QSL compatible with lower lattice symmetry to emerge. These questions are however well beyond the scope of present work and will be addressed in future. 

Such a model at intermediate values of $\delta$ may be possible to realize in experiments by careful substitution of either ligands or non-magnetic ions within alternate square plaquettes of square lattice Mott insulators. There are recent studies where such substitution has been done, albeit in a random manner \cite{Mustonen18, Hong21}. This leads to an intermediate random-singlet phase. However, one strategy may be to consider a square lattice material which is non-magnetic with two different atoms on alternating sites. 
If a magnetic material with a similar lattice constant is grown or stacked (with appropriate displacement) on such a substrate there is a possibility to realize the model discussed in this paper. It may sound wishful thinking, but with current developments in the area of 2d materials it may not be unimaginable. 

Our work opens further questions and directions which will be pursued in future.  To mention a few, one immediate question could be to study this model in an external field. More generally, our work motivates further investigations of quantum phase transitions out of an altermagnet. We expect our work to motivate detailed numerical studies of this model, particularly to investigate the quantum disordered phase in the pursuit of quantum spin liquid.  

\section{Acknowledgements}

We acknowledge the support of the Department of Atomic Energy, Government of India, under Project Identification No. RTI 40007.

\appendix
\section{N\'{e}el/Altermagnet phase}
\label{sec:appN}

In this appendix quote the next order terms that are relevant for $1/S$ corrections in the AM phase.
\begin{widetext}
\begin{align}
\frac{H_4}{J/4} &= -\sum_{\substack{\langle i,j \rangle \\ i \in A, j \in B}}( 4\ad_i\ai_i \bd_j\bi_j + \ad_i\ad_i \ai_i\bd_j 
+ \ad_i\bd_j \bd_j\bi_j + \ai_i\bd_j \bi_j\bi_j + \ad_i\ai_i \ai_i\bi_j) \nonumber\\
& + \lambda (1+\delta)\sum_{\substack{i\in A\\j=i+\delta_{xy}}}(4\ad_i\ai_i \ad_j\ai_j - \ad_i\ad_j \ai_j\ai_j 
- \ad_i\ad_i \ai_i\ai_j - \ai_i\ad_j \ad_j\ai_j - \ad_i\ai_i \ai_i\ad_j) \nonumber\\
& + \lambda (1-\delta)\sum_{\substack{i\in A\\j=i+\delta'_{xy}}}(4\ad_i\ai_i \ad_j\ai_j - \ad_i\ad_j \ai_j\ai_j 
- \ad_i\ad_i \ai_i\ai_j - \ai_i\ad_j \ad_j\ai_j - \ad_i\ai_i \ai_i\ad_j) \nonumber\\
& + \lambda (1+\delta)\sum_{\substack{i\in B\\j=i+\delta'_{xy}}}(4\bd_i\bi_i \bd_j\bi_j - \bd_i\bd_j \bi_j\bi_j 
- \bd_i\bd_i \bi_i\bi_j - \bi_i\bd_j \bd_j\bi_j - \bd_i\bi_i \bi_i\bd_j) \nonumber\\
& + \lambda (1-\delta)\sum_{\substack{i\in B\\j=i+\delta_{xy}}}(4\bd_i\bi_i \bd_j\bi_j - \bd_i\bd_j \bi_j\bi_j 
- \bd_i\bd_i \bi_i\bi_j - \bi_i\bd_j \bd_j\bi_j - \bd_i\bi_i \bi_i\bd_j)
\end{align}
\end{widetext}

\section{CAF phase}
\label{sec:appC}

In this appendix we briefly discuss the details of calculations related to the Bosonic Bogoliubov transformation in the CAF phase. Recall Eq. (\ref{eq:h2ck}) - (\ref{eq:hkc}) in the main text. To obtain the magnon modes and other observables we have to perform a Bosonic Bogoliubov transformation such that
\begin{equation}
T^{\dagger} h_{\kk} T = \Omega_{\kk} \,,   
\end{equation}
where $\Omega_{\kk} = diag(\omega_{1\kk}, \omega_{2\kk},\omega_{1\kk}, \omega_{2\kk})$ is the diagonal matrix containing the magnon mode dispersions. The matrix $T$ is an anti-unitary matrix containing the Bogoliubov coefficients, such that 
\begin{equation}
T^{\dagger} \Sigma T = \Sigma \,,   
\end{equation}
where $\Sigma = diag(1,1,-1,-1)$. The coefficients, $t_{ij}$, appearing in Eqs. (\ref{eq:opC1})-(\ref{eq:opC4}) and Eqs. (\ref{eq:z1c})-(\ref{eq:z2c}) are the matrix elements of $T$.  Additionally, $\psi_{\kk} = T \phi_{\kk}$, where $\psi_{\kk}$ is the original magnon operator basis (see above Eq. (\ref{eq:hkc})) and $\phi_{\kk}$ is the basis in which the leading order bilinear Hamiltonian is diagonal. Physical observables such as the order parameter, spin structure factor etc. can be now expressed in terms of the coefficients of the $T$ matrix, which are the Bogoliubov coefficients. In general, these are to be evaluated numerically at a given $\kk$ point.

In addition, we also quote the Hamiltonian at the next order in $1/S$ expansion relevant to evaluate the next order corrections.
\begin{widetext}
\begin{align}
\frac{H_4}{J/4} &= \sum_{\substack{\langle i,j\rangle\\ i \in A ,j \in D}}(4\ad_i\ai_i\dd_j\di_j - \ai_i\dd_j\dd_j\di_j 
- \ad_i\ai_i\ai_i\dd_j - \ad_i\dd_j\di_j\di_j - \ad_i\ad_i\ai_i\di_j) \nonumber\\
& +\sum_{\substack{\langle i,j\rangle\\ i \in C, j \in B}}(4\cd_i\ci_i\bd_j\bi_j - \ci_i\bd_j\bd_j\bi_j 
- \cd_i\ci_i\ci_i\bd_j - \cd_i\bd_j\bi_j\bi_j - \cd_i\cd_i\ci_i\bi_j) \nonumber\\
& -\sum_{\substack{\langle i,j\rangle\\ i \in A, j \in B}}(4\ad_i\ai_i\bd_j\bi_j + \ad_i\ai_i\ai_i\bi_j 
+ \ai_i\bd_j\bi_j\bi_j + \ad_i\bd_j\bd_j\bi_j + \ad_i\ad_i\ai_i\bd_j) \nonumber\\
& -\sum_{\substack{\langle i,j\rangle\\ i \in C, j \in D}}(4\cd_i\ci_i\dd_j\di_j + \cd_i\ci_i\ci_i\di_j 
+ \ci_i\dd_j\di_j\di_j + \cd_i\dd_j\dd_j\di_j + \cd_i\cd_i\ci_i\dd_j) \nonumber\\
& -\lambda (1+\delta)\sum_{\substack{i \in A\\j=\pm \delta_{xy}}}(4\ad_i\ai_i\cd_j\ci_j + \ad_i\ai_i\ai_i\ci_j 
+ \ai_i\cd_j\ci_j\ci_j + \ad_i\cd_j\cd_j\ci_j + \ad_i\ad_i\ai_i\cd_j) \nonumber\\
& -\lambda (1-\delta)\sum_{\substack{i \in A\\j=\pm \delta'_{xy}}}(4\ad_i\ai_i\cd_j\ci_j + \ad_i\ai_i\ai_i\ci_j 
+ \ai_i\cd_j\ci_j\ci_j + \ad_i\cd_j\cd_j\ci_j + \ad_i\ad_i\ai_i\cd_j) \nonumber\\
& -\lambda (1+\delta)\sum_{\substack{i \in B\\j=\pm \delta'_{xy}}}(4\bd_i\bi_i\dd_j\di_j + \bd_i\bi_i\bi_i\di_j 
+ \bi_i\dd_j\di_j\di_j + \bd_i\dd_j\dd_j\di_j + \bd_i\bd_i\bi_i\dd_j) \nonumber\\
& -\lambda (1-\delta)\sum_{\substack{i \in B\\j=\pm \delta_{xy}}}(4\bd_i\bi_i\dd_j\di_j + \bd_i\bi_i\bi_i\di_j 
+ \bi_i\dd_j\di_j\di_j + \bd_i\dd_j\dd_j\di_j + \bd_i\bd_i\bi_i\dd_j) \,.
\end{align}
\end{widetext}

\bibliography{swt}

\end{document}